\documentclass[10pt,conference]{IEEEtran}
\usepackage{cite}
\usepackage{amsmath,amssymb,amsfonts,amsthm}
\usepackage{mathtools}
\usepackage{mathrsfs}
\usepackage{algorithm}
\usepackage{algorithmicx}
\usepackage[noend]{algcompatible}
\usepackage{graphicx}
\usepackage{textcomp}
\usepackage{xcolor}
\usepackage{dashbox}
\usepackage{mathtools}
\usepackage[leftmargin=3em]{quoting}
\usepackage{graphicx}
\usepackage{mathtools}
\usepackage{mathrsfs}
\usepackage{hyperref}
\hypersetup{
  colorlinks   = true,    
  urlcolor     = black,    
  linkcolor    = black,    
  citecolor    = black      
}
\def\BibTeX{{\rm B\kern-.05em{\sc i\kern-.025em b}\kern-.08em
    T\kern-.1667em\lower.7ex\hbox{E}\kern-.125emX}}

\theoremstyle{definition}

\newtheorem{definition}{Definition}
\newtheorem{theorem}{Theorem}
\newtheorem{lemma}{Lemma}
\newtheorem{corollary}{Corollary}

\newcommand\NoDo{\renewcommand\algorithmicdo{}}

\newcommand\NoThen{\renewcommand\algorithmicthen{}}

\algnewcommand\algorithmicto{\textbf{to}}
\algnewcommand\RETURN{\State \textbf{return} }

\DeclarePairedDelimiter\ceil{\lceil}{\rceil}

\usepackage{lipsum}

\newcommand\blfootnote[1]{%
  \begingroup
  \renewcommand\thefootnote{}\footnote{#1}%
  \addtocounter{footnote}{-1}%
  \endgroup
}

\begin{document}
    \title{
        On the Minimal Knowledge Required for Solving Stellar Consensus
    }

    \author{
    \IEEEauthorblockN{
        Robin Vassantlal,
        Hasan Heydari, and
        Alysson Bessani} 

        \IEEEauthorblockA{LASIGE, Faculdade de Ciências, Universidade de Lisboa, Portugal}
    
        \{rvassantlal, hheydari, anbessani\}@ciencias.ulisboa.pt
    }
    
    \maketitle
    

    \begin{abstract}
        Byzantine Consensus is fundamental for building consistent and fault-tolerant distributed systems. 
In traditional quorum-based consensus protocols, quorums are defined using globally known assumptions shared among all participants. 
Motivated by decentralized applications on open networks, the Stellar blockchain relaxes these global assumptions by allowing each participant to define its quorums using local information. 
A similar model called Consensus with Unknown Participants (CUP) studies the minimal knowledge required to solve consensus in ad-hoc networks where each participant knows only a subset of other participants of the system.
We prove that Stellar cannot solve consensus using the initial knowledge provided to participants in the CUP model, even though CUP can.
We propose an oracle called \emph{sink detector} that augments this knowledge, enabling Stellar participants to solve consensus.

    \end{abstract}


\blfootnote{This is a preprint of a paper to appear at the 43rd IEEE International Conference on Distributed Computing Systems (ICDCS 2023).}

    \section{Introduction}

Consensus is a fundamental building block for distributed systems that remain available and consistent despite the failure of some participants~\cite{lamport_1982,schneider_1990,castro_1999}. 
In this problem, participating processes agree on a common value from the initially proposed values. 
This problem was extensively studied considering different synchrony assumptions (e.g., partially synchronous) and failure models (e.g., Byzantine failures) in the permissioned setting, where the set of participants and the fault threshold is known \textit{a priori} by all participants (e.g., ~\cite{castro_1999,dwork_1988,lamport_1998,ongaro_2014}). 

The Nakamoto consensus protocol~\cite{nakamoto_2008}, used in Bitcoin, makes it possible to solve consensus without having a single global view of the system. 
In Nakamoto consensus, the set of all participants is unknown, and the system's fault tolerance is determined based on the total amount of computing power controlled by the adversary. 
Even though this protocol opens doors for anyone to participate in consensus and is scalable in the number of participants, its performance is orders of magnitude lower than consensus protocols for the permissioned setting~\cite{vukolic_2015}.

However, with the popularization of blockchains, the demand for scaling consensus to many participants while maintaining high performance led researchers to propose interesting alternatives. 
Examples include hybrid consensus~\cite{abraham_2016,decker_2016,pass_2017_hybrid_consensus} and asymmetric trust-based protocols~\cite{garcia_2018,lokhava_2019,schawartz_2014}. 
In the former, a committee of participants is randomly selected from a network of unknown size in proportion to the resources they control (e.g., computing power or stake) to execute a traditional permissioned Byzantine fault-tolerant consensus (e.g., PBFT~\cite{castro_1999}). 
In the latter, the global knowledge about the system membership and fault threshold required in the permissioned consensus is relaxed by enabling each participant to declare a partial view of the participants it can trust.
This paper focuses on one of the most well-known asymmetric trust-based protocols called Stellar~\cite{lokhava_2019,mazieres_2015}.

The Stellar blockchain enables exchanging digital assets worldwide without relying on centralized authorities such as banks. 
Stellar comprises two main elements: a Byzantine quorum-based consensus protocol called SCP (Stellar Consensus Protocol) and a network of trust called the Stellar network.
SCP maintains a consistent ledger of transactions where participants neither need to know all participants nor the maximum number of participants that can fail in partially synchronous systems.
Besides, it allows anyone to join the network without reconfiguring the system. 

SCP is executed on the Stellar network built using trust relationships declared by each participant.
More specifically, at the beginning of the execution, each participant in Stellar has only access to a set of slices, where each slice is a set of participants.
Even though it is unclear how slices are defined in Stellar~\cite{mazieres_2015}, in practice, these slices are manually defined based on a list of trusted participants. 
The combination of these slices forms quorums.
SCP can solve consensus in the Stellar network if quorums satisfy a property called consensus cluster~\cite{losa_2019}.

\begin{figure}[t!]
    \begin{minipage}{0.56\linewidth}
        \includegraphics[scale=0.22]{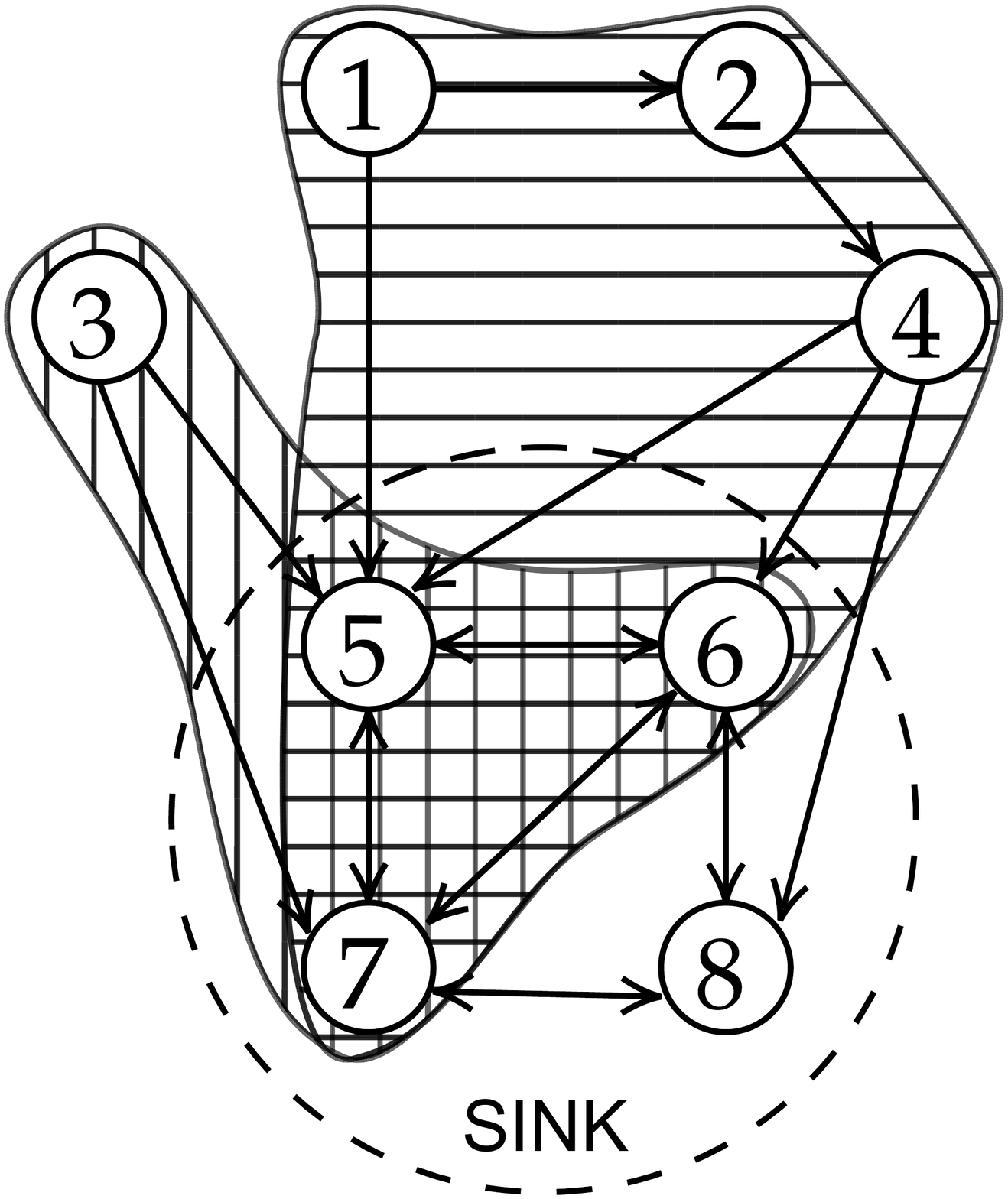}
    \end{minipage}
    \hspace{-0.2cm}
    \begin{minipage}{0.43\linewidth}
        \textbf{Participants' knowledge:}\\
        $\mathit{PD}_1 = \{ 2,5 \}$ \\
        $\mathit{PD}_2 = \{ 4 \}$ \\
        $\mathit{PD}_3 = \{ 5,7 \}$ \\
        $\mathit{PD}_4 = \{ 5,6,8 \}$ \\
        \dbox{
        \hspace{-1em}
        \parbox{0.7\linewidth}{
            $\mathit{PD}_5 = \{ 6,7 \}$ \\
            $\mathit{PD}_6 = \{ 5,7,8 \}$ \\
            $\mathit{PD}_7 = \{ 5,6,8 \}$ \\
            $\mathit{PD}_8 = \{ 6,7 \}$
        }
        } 
    \end{minipage}
        \caption{
        An example of a knowledge connectivity graph with 8 participants. 
        For each participant $i \in \{ 1,2,\dots,8 \}$, $\mathit{PD}_i$ shows the information provided by its participant detector, i.e., the knowledge of $i$. 
        Participants 5, 6, 7, and 8 form the sink component.
    }
    \label{fig:cup}
\end{figure}

There is a similar line of research on a model called Consensus with Unknown Participants (CUP) that studies the knowledge required to solve consensus in settings in which each participant joins the network knowing only a subset of other participants and the fault threshold of the system~\cite{alchieri_2018,cavin_2004,cavin_2005,greve_2007}.
In this model, each participant's knowledge about the existence of other participants is encapsulated in a local oracle called \emph{participant detector}. 
The union of the information the participant detectors provide forms a knowledge connectivity graph, where a vertex is a participant, and an edge represents the knowledge between two participants. 
For example, in Fig.~\ref{fig:cup}, participant 1 initially knows participants 2 and 5.

The CUP model allows the establishment of the minimal knowledge necessary and sufficient under specific synchrony and fault assumptions to solve consensus.
The knowledge requirement increases as the synchrony assumptions are relaxed and the fault assumptions get stronger.
For example, each participant in Byzantine Fault-Tolerant (BFT) CUP~\cite{alchieri_2018} requires more knowledge than in the fault-free CUP model~\cite{cavin_2004}.

The main question we address in this paper is to determine whether SCP can be executed with the minimal knowledge established in the CUP model, i.e., can each participant define its slices using only a list of participants and a fault threshold?
We present two attempts to answer this question. 
The first attempt locally defines slices for each participant using only a list provided by its participant detector and the fault threshold.
We prove that this information is not enough for SCP.
The second attempt successfully defines slices using some extra information obtained by increasing the knowledge of each participant.
We indeed define and implement an oracle called \emph{sink detector} that provides such required extra information.
These results show that, differently from the BFT-CUP protocol~\cite{alchieri_2018}, SCP is not powerful enough to solve consensus under minimal proven initial knowledge conditions; however, this limitation can be circumvented using a sink detector.

\vspace{1em}
\noindent\textbf{Contributions.}
This paper makes the following contributions:
\begin{itemize}   
    \item We show that SCP cannot solve consensus when each participant has only the minimum knowledge required to solve consensus.

    \item We propose an oracle -- \emph{sink detector} -- by which participants can solve consensus using SCP when each participant has access only to the same knowledge required to solve CUP.
\end{itemize}

\vspace{1em}
\noindent\textbf{Paper organization.} 
The remainder of the paper is organized as follows. 
Section~\ref{sec:related-work} presents the related work.
Section~\ref{sec:preliminaries} presents our system model and describes the background of this paper. 
Section~\ref{sec:requires-more} shows that SCP cannot solve consensus when each participant is given the same knowledge required by CUP. 
Section~\ref{sec:defining-slices} shows how slices can be defined using a participant detector, the fault threshold, and a sink detector. 
Section~\ref{sec:implementing-sink} presents the implementation of the sink detector.
Finally, Section~\ref{sec:conclusion} concludes the paper.

    \section{Related Work}
\label{sec:related-work}

\noindent\textbf{Stellar.} 
The design of protocols for participants with different trust assumptions (i.e., participants can trust different sets of participants) was first studied in~\cite{damgard_2007}.
Ripple~\cite{schawartz_2014,chase_2018} attempts to use this approach to solve consensus in the permissionless setting, with the goal of establishing an efficient blockchain infrastructure; however, the goal is not completely attained due to existing safety and liveness violations~\cite{amores_2020}.
Stellar~\cite{mazieres_2015,lokhava_2019}, which is based on the Federated Byzantine Quorum System (FBQS) (formally studied afterward in \cite{garcia_2018}), successfully achieves that goal. 
In this new attempt, a network of trust emerges from the partial view declared by each participant.
Solving consensus in this network is guaranteed if it satisfies the \emph{intact set} property, which states that all correct participants must form a quorum, and any two quorums formed by correct participants must intersect.

A connection between FBQS and the dissemination Byzantine quorum systems~\cite{malkhi_1998} was established in \cite{garcia_2018} and \cite{garcia_2019}, demonstrating how to construct a dissemination Byzantine quorum system that corresponds to an FBQS.
Later work by Losa et al.~\cite{losa_2019} generalized FBQS to Personal Byzantine Quorum System (PBQS).
They proved that solving consensus with weaker properties than the intact set is possible using a \emph{consensus cluster}.
Specifically, forming a quorum by all correct participants is not required in the consensus cluster.
Cachin and Tackmann~\cite{cachin_2019} extended Byzantine Quorum Systems (BQS)~\cite{malkhi_1998} from the symmetric trust model to the asymmetric model, which made it easier to compare PBQS with the classical BQS model.
Recently, Cachin et al. generalized the asymmetric trust model by allowing each participant to make assumptions about the failures of participants it knows and, through transitivity, about failures of participants indirectly known by it~\cite{cachin_2022}.

\vspace{1em}
\noindent\textbf{Consensus with Unknown Participants (CUP).}
The Consensus with Unknown Participants (CUP) problem has been developed through various steps, adapting to different system models.
Initially, the problem was defined by Cavin et al. \cite{cavin_2004} for failure-free asynchronous systems by introducing the participant detector abstraction to provide participants with initial information about the system membership.
The information provided to participants collectively forms a knowledge connectivity graph.
This work determines a knowledge connectivity graph's necessary and sufficient properties to solve CUP.
Later, the CUP problem was solved in~\cite{cavin_2005} for crash-prone systems enriched with the Perfect ($\mathcal{P}$) failure detector \cite{chandra_1996}. 
As $\mathcal{P}$ requires synchrony, Greve and Tixeuil~\cite{greve_2007} relaxed the assumption to partial synchrony~\cite{dwork_1988} by increasing the minimum required knowledge, i.e., increasing the connections in the knowledge connectivity graph (detailed in the next section), which they show to be minimum to tolerate crash failures without requiring synchrony.
The last milestone expanded CUP to tolerate Byzantine failures by introducing the BFT-CUP protocol~\cite{alchieri_2008, alchieri_2018}.

More recently, it was shown that synchrony is required to solve consensus without knowing the number of participants and the fault threshold~\cite{khanchandani_2021}.

\vspace{1em}
\noindent\textbf{Sleepy model.} Both Stellar and CUP solve consensus in partially synchronous systems, where participants can be correct or faulty.
It is also assumed that correct members participate throughout the whole execution of the protocols, which is unrealistic in practice.
In the recently proposed sleepy model~\cite{momose_2022,pass_2017}, participants in a synchronous system are assumed to be either awake or asleep, whereas awake participants can be faulty or correct.
In this model, the system's fault tolerance dynamically changes as the participants transition between awake and asleep states.
Further, consensus can be solved if the majority of awake participants are correct at any time.
Differently from Stellar and CUP, in this model, all participants know the set of participants in the system.

\vspace{1em}
\noindent\textbf{Consensus in directed graphs.}
Somewhat similar to CUP, several works study consensus in directed graphs, e.g., ~\cite{biely_2012,vaidya_2012,biely_2018,tseng_2015}.
However, those works study the requirements of the underlying communication graph to solve consensus under different assumptions.
For example, Tseng and Vaidya~\cite{tseng_2015} proved the minimal conditions of the underlying communication graph, where a participant $i$ can send messages to participant $j$ if there is a directed edge from $i$ to $j$ in the graph; otherwise, $i$ cannot send messages to $j$.
Typically, these works assume the set of participants, and the underlying communication graph is known by all participants.
In the CUP model considered in this paper, the communication graph is complete, and the goal is to study the required initial knowledge about other participants, which forms a knowledge connectivity graph (see Fig.~\ref{fig:cup}), to solve consensus without knowing the total number of participants in the system.

    \section{Preliminaries}\label{sec:preliminaries}

\subsection{System Model}\label{sec:system-model}
We consider a partially synchronous distributed system~\cite{dwork_1988} in which the network and processes (also called participants) may behave asynchronously until some \emph{unknown} global stabilization time GST after which the system becomes synchronous, with \textit{unknown time bounds for computation and communication}. 
This system is composed of a finite set $\Pi$ of processes drawn from a larger universe $\mathcal{U}$.
In a known network, $\Pi$ is known to every process.
In contrast, in an unknown network, process $i \in \Pi$ knows only a subset $\Pi_{i} \subseteq \Pi$.
We assume that $\Pi$ is static during the execution of an instance of consensus, i.e., no process leaves or joins the system. 
Our analysis is for a single instance of consensus.

Processes are subject to Byzantine failures~\cite{lamport_1982}. 
A process that does not follow its algorithm is called \textit{faulty}. 
A process that is not faulty is said to be \textit{correct}. 
During the execution of an instance of consensus, we denote $W$ as the set of processes that remain correct and define $F=\Pi \setminus W$ as the set of processes that can fail.
We consider a \emph{static Byzantine adversary}, i.e., $F$ is fixed at the beginning of the protocol execution by the adversary.
Even though $W$ and $F$ are unknown, $|F| \leq f$, where $f$ is the maximum number of faulty processes.
We assume that $f$ is known unless stated otherwise. 

We further assume that all processes have a unique id, and it is infeasible for a faulty process to obtain additional ids to launch a \emph{Sybil attack} \cite{douceur_2002}.
Processes communicate by message passing through authenticated 
 and reliable point-to-point channels. 
A process $i$ may only send a message directly to another process $j$ if $j \in \Pi_{i}$, i.e., if $i$ knows $j$. 
Of course if $i$ sends a message to $j$ such that $i \notin \Pi_{j}$, upon receipt of the message, $j$ may add $i$ to $\Pi_{j}$, i.e., $j$ now knows $i$ and can send messages to it. 


\subsection{The Consensus Problem} 
In the consensus problem, each process $i$ \emph{proposes} a value $v_i$, and all correct processes \emph{decide} the same value $v$ among the proposed values. 
Formally, any protocol that solves consensus must satisfy the following properties (e.g.,~\cite{fischer_1983}):
\begin{itemize}
    \item \emph{Validity:} a correct process decides $v$, then $v$ was proposed by some process.
    \item \emph{Agreement:} no two correct processes decide differently.
    \item \emph{Termination:} every correct process eventually decides some value.
    \item \emph{Integrity:} every correct process decides at most once.
\end{itemize}

\subsection{Byzantine Quorum Systems}
Byzantine Quorum Systems (BQS) enable solving consensus despite Byzantine failures~\cite{malkhi_1997}. 
A BQS is composed of a set of quorums $\mathcal{Q}$, where each quorum $Q \in \mathcal{Q}$ is a subset of processes that satisfies two properties:

\begin{itemize}
    \item \emph{Consistency}: every two quorums intersect in at least one correct process, i.e., $\forall Q_1, Q_2 \in \mathcal{Q} : Q_1 \cap Q_2 \cap W \neq \emptyset$.
    \item \emph{Availability}: there is at least one quorum composed only by correct processes, i.e., $\exists Q \in \mathcal{Q} : Q \subseteq W$.
\end{itemize}

If one of these two properties is not satisfied, the correctness of a consensus protocol based on BQS cannot be guaranteed.
Furthermore, in BQS, $\Pi$ and $f$ are known for every process.
Notice that quorums in a BQS are shared among processes, i.e., if a set of processes $Q \subseteq \Pi$ is a quorum in a BQS, then $Q$ is a quorum for every process in $\Pi$.

\subsection{Stellar Model}
The Stellar model relaxes the global knowledge assumption about $\Pi$ and $f$ by employing Federated Byzantine Quorum System (FBQS)~\cite{garcia_2018,mazieres_2015,lokhava_2019}.
In FBQS, at the beginning of the execution, each process $i$ has only access to its quorum \emph{slices}, which are simply referred to as slices.
Each slice of a process $i$ is a set of processes that $i$ trusts.
We denote the set of all slices of a process $i$ by $\mathcal{S}_i$. 
Given a set $A \subseteq \Pi$, we define $\mathcal{S}_{A} = \{ \mathcal{S}_i \ | \ \forall i \in A \}$, and for each $i \in A$, $\mathcal{S}_{A}[i]$ equals $\mathcal{S}_i$. 
We consider that the union of all slices of $i$ is $\Pi_i$, i.e., $\bigcup_{S \in \mathcal{S}_i} S = \Pi_i$. 

\begin{definition}[Quorum]\label{def:quorum}
A set of processes $Q$ is a quorum if each process $i \in Q$ has at least a slice contained within $Q$, i.e., $\forall i \in Q, \exists S \in \mathcal{S}_i : S \subseteq Q$.
\end{definition}

We say that quorum $Q$ is a quorum for a process $i$ if $i$ belongs to it and it contains at least a slice of $i$.
We denote the set of all quorums of a process $i$ by $\mathcal{Q}_{i}$.
Each process $i$ attaches $\mathcal{S}_i$ to all of the messages it sends so that any other process knows $\mathcal{S}_i$ by receiving a message from $i$.
We introduce a function $\mathtt{is\_quorum}(Q,\mathcal{S}_Q)$ (Algorithm~\ref{alg:is_quorum}) by which a process $i$ can identify whether $Q$ is a quorum using $\mathcal{S}_Q=\{\mathcal{S}_j \ | \ \forall j \in Q\}$.

\begin{algorithm}[t!]
    \caption{Determining if a set $Q$ is a quorum.}
    \label{alg:is_quorum}
    \begin{algorithmic}[1]
        \STATEx {\hspace{-1.67em}\textbf{function} $\mathtt{is\_quorum}(Q, \mathcal{S}_Q)$}
            \NoDo
            \FOR{all $i \in Q$}            
                \NoThen
                \IF{$\nexists S \in \mathcal{S}_{Q}[i] : S \subseteq Q$}
                    \RETURN{$\mathit{false}$}
                \ENDIF
            \ENDFOR
            \RETURN{$\mathit{true}$}
    \end{algorithmic}
\end{algorithm}

In the Stellar model, correct processes can solve consensus under partial synchrony if quorums form a \emph{consensus cluster} (Definition~\ref{def:consensus_cluster}).
A consensus cluster emerges if quorums are \emph{intertwined} as defined below.
The following three definitions are adapted from \cite{losa_2019}.

\begin{definition}[Intertwined]\label{def:intertwined}
    A set $I$ of correct processes is intertwined if, for any two members $i$ and $j$ of $I$, the intersection of any quorum $Q$ of $i$ and any quorum $Q'$ of $j$ contains at least one correct process, i.e., 
    $\forall i,j \in I, \forall Q \in \mathcal{Q}_i, \forall Q' \in \mathcal{Q}_j : Q \cap Q' \cap W \neq \emptyset$.
\end{definition}

\begin{definition}[Consensus cluster]
    \label{def:consensus_cluster}
    A subset $I \subseteq W$ of the correct processes is a consensus cluster when:
    \begin{itemize}
        \item \emph{Quorum Intersection}: $I$ is intertwined, and
        \item \emph{Quorum Availability}: if $i \in I$ then there is a quorum $Q$ of $i$ such that every member of $Q$ is correct and is inside $I$, i.e., $Q \subseteq I$.
    \end{itemize}
\end{definition}

The quorum intersection property allows to guarantee the agreement and integrity property of consensus, while quorum availability ensures that every correct process makes progress by having at least a quorum composed entirely of correct processes, i.e., it enforces the termination property of consensus.

\begin{definition}[Maximal consensus cluster]
    A maximal consensus cluster is a consensus cluster that is not a strict subset of any other consensus cluster. 
\end{definition}

All correct processes of the system can solve consensus if there is exactly one maximal consensus cluster $C$ in the system such that $C = W$~\cite{losa_2019}.
To see an example of slices and quorums, consider the graph depicted in Fig.~\ref{fig:cup}.
In this graph, we assume that $W=\{1,2,\dots,7\}$ and $F=\{8\}$. 
Besides, let slices of each correct process be defined as follows:
$\mathcal{S}_1 = \{\{2, 5\}\}$,
$\mathcal{S}_2 = \{\{4\}\}$,
$\mathcal{S}_3 = \{\{5, 7\}\}$,
$\mathcal{S}_4 = \{\{5, 6\}, \{6, 8\}\}$,
$\mathcal{S}_5 = \{\{6, 7\}\}$,
$\mathcal{S}_6 = \{\{5, 7\}, \{7, 8\}\}$,
$\mathcal{S}_7 = \{\{5, 6\}, \{6, 8\}\}$.
Since a Byzantine process can define its slices arbitrarily, it is not required to define its slices; however, correct processes must define their slices so that the maximal consensus cluster can emerge.
Notice that, with these slices, there is a quorum for each correct process (e.g., $1$'s quorum is the area with horizontal lines, and $3$'s quorum is the area with vertical lines).
Since all those quorums intersect at quorums of $5$, $6$, and $7$ (i.e., $Q_5=Q_6=Q_7=\{5,6,7\}$ \---- the area with squares), which are composed of correct processes, every two correct processes are intertwined.
In this example, there are a few consensus clusters, such as $C_1 = \{5,6,7\}$ and $C_2 = \{1,2,\dots,7\}$, but $C_2$ is the only maximal consensus cluster.

\subsection{CUP Model}
The Consensus with Unknown Participants (CUP) model~\cite{cavin_2004} solves consensus in a distributed system where processes' knowledge about the system composition is incomplete. 
This model is useful for studying the necessary and sufficient knowledge conditions that processes require in order to solve consensus under different assumptions.

In CUP, the knowledge connectivity is encapsulated in an oracle called a \emph{Participant Detector} (PD).
A PD can be seen as a distributed oracle that provides each process hints about the participating processes in the distributed computation. 
Let $\mathit{PD}_i$ be defined as the participant detector of a process $i$, such that $\mathit{PD}_i$ returns a set of processes $\Pi_{i} \subseteq \Pi$ to which $i$ can initially contact.
We say a process $j$ is a neighbor of another process $i$ if and only if $j \in \mathit{PD}_i$.
The information provided by the participant detectors of all processes forms a \emph{knowledge connectivity graph} (see definition below), which is a directed graph since the initial knowledge provided by different PDs is not necessarily bidirectional, i.e., $i$ knows $j$, but $j$ might not know $i$.

\begin{definition}[Knowledge connectivity graph \cite{alchieri_2018}]
\label{def:knowledge_connectivity_graph}
    Let $G_{\mathit{di}} = (V_{\mathit{di}}, E_{\mathit{di}})$ be the directed graph representing the knowledge relation determined by the PD oracle. Then, $V_{\mathit{di}} = \Pi$ and $(i, j) \in E_{\mathit{di}}$ if and only if $j \in \mathit{PD}_i$, i.e., $i$ knows $j$.
\end{definition}

It is important to remark that the knowledge connectivity graph defines the list of processes that every process initially knows in the system, \emph{not their network's connectivity}.
In CUP, at the beginning of the execution, each process $i$ has only access to $\mathit{PD}_i$ and $f$.
Access to PD is required since processes cannot solve any nontrivial distributed coordination task without having some initial knowledge about other processes~\cite{alchieri_2018}.
Furthermore, when $n$ is unknown, processes cannot solve consensus in non-synchronous systems without knowing $f$~\cite{khanchandani_2021}.

The undirected graph obtained from the directed knowledge connectivity graph $G_{\mathit{di}} = (V_{\mathit{di}}, E_{\mathit{di}})$  is defined as $G = (V_{\mathit{di}}, \{(i, j) : (i, j) \in E_{\mathit{di}} \lor (j, i) \in E_{\mathit{di}}\})$. 
A component $G_{\mathit{sink}} = (V_{\mathit{sink}}, E_{\mathit{sink}})$ of $G_{\mathit{di}}$ is a sink component if and only if there is no path from a node in $G_{\mathit{sink}}$ to other nodes of $G_{\mathit{di}}$, except nodes in $G_{\mathit{sink}}$ itself. 
A process $i \in V_{\mathit{di}}$ is a \emph{sink member} if and only if $i \in V_{\mathit{sink}}$; otherwise, $i$ is a \emph{non-sink member}. 
See Fig.~\ref{fig:cup} for an example.

The Byzantine Fault-Tolerant (BFT) CUP problem can be solved under partial synchrony if the knowledge connectivity graph of processes satisfies the $k$-One Sink Reducibility property~\cite{alchieri_2018,alchieri_2008}.
This property ensures that every process can reach the sink members, and every correct sink member can discover the whole sink.
As soon as the sink is discovered, sink members solve consensus among themselves by executing a consensus protocol (e.g., PBFT~\cite{castro_1999}).
Then, they disseminate the decided value to non-sink members. 
Notice that having multiple sinks might violate the agreement property of consensus because each sink might remain unaware of other sinks until deciding some value, yielding the possibility of deciding distinct values.

\begin{definition}[$k-$One Sink Reducibility (OSR) PD~\cite{alchieri_2018}]
    \label{def:osr}
    This class of PD contains all knowledge connectivity graphs $G_{\mathit{di}}$ such that:
    \begin{enumerate}
        \item the undirected graph $G$ obtained from $G_{\mathit{di}}$ is connected;
        \item the directed acyclic graph obtained by reducing $G_{\mathit{di}}$ to its strongly connected components has exactly one sink, namely $G_\mathit{sink}$;
        \item the sink component $G_\mathit{sink}$ is $k-$strongly connected;\footnote{
        A graph $G$ is said to be $k$-strongly connected if, for any pair $(i,j)$ of nodes in $G$, $i$ can reach $j$ through at least $k$ node-disjoint paths in $G$.}
        \item for all $i,j \in V_{\mathit{di}}$, such that $i \notin G_\mathit{sink}$ and $j \in G_\mathit{sink}$, there are at least $k$ node-disjoint paths from $i$ to $j$.
    \end{enumerate}
\end{definition}

The \emph{safe Byzantine failure pattern} defines the parameter $k$ of $k$-OSR PD by considering the location of up to $f$ failures in $G_{\mathit{di}}$.

\begin{definition}[Safe Byzantine failure pattern~\cite{alchieri_2018}]
    \label{def:safe_byzatine_failure_pattern}
    Let $G_{\mathit{di}}$ be a knowledge connectivity graph, $f$ be the maximum number of processes in $G_{\mathit{di}}$ that may fail, and $F$ be the set of faulty processes in $G_{\mathit{di}}$ during an execution. The safe Byzantine failure pattern for $G_{\mathit{di}}$ and $F$ is the graph $G_{\mathit{safe}} = G_{\mathit{di}} \setminus F : (F \subset G_{\mathit{di}}) \wedge (|F| \leq f) \wedge (G_{\mathit{di}} \setminus F \in (f+1)-$OSR$)$.
\end{definition}

If the safe Byzantine failure pattern holds during the execution of consensus in $G_{\mathit{di}}$, then we say that $G_{\mathit{di}}$ is \emph{Byzantine-safe} for $F$.
The following theorem from~\cite{alchieri_2018} determines the minimal requirements to solve consensus in the BFT-CUP model.

\begin{theorem}\label{thm:consensus_conditions_in_cup}
     Consensus is solvable in the BFT-CUP model if there is a knowledge connectivity graph that is Byzantine-safe for $F$, and its sink component contains at least $2f+1$ correct processes.
\end{theorem}

Throughout the paper, when we use $\mathit{PD}_i$, where $i$ is a process, we assume that the union of $\mathit{PD}_1, \mathit{PD}_2, \dots, \mathit{PD}_{|\Pi|}$ forms a $k$-OSR graph that is Byzantine-safe for $F$ and its sink component has at least $2f+1$ correct processes.

\subsection{Threshold-based Analysis}
Recall that both the Stellar and BFT-CUP models solve consensus under partial synchrony.
Since it is proved that solving consensus without knowing $n$ and $f$ (or a fail-prone system in general) is impossible in non-synchronous systems~\cite{khanchandani_2021}, both Stellar and BFT-CUP require knowledge of $f$ or a fail-prone system.
However, only BFT-CUP explicitly considers $f$.

Since the main objective of this paper is to analyze the knowledge requirement of Stellar and compare it with the BFT-CUP model, we consider threshold-based systems for the Stellar model to have a common ground and also for simplicity, i.e., we use $f$ to define slices and quorums. 
Consequently, in the remaining part of the paper, we say that 
a set $I$ of correct processes is intertwined if any two quorums $Q$ and $Q'$ of members of $I$ satisfy $|Q \cap Q'| > f$.

    \section{Defining Slices in the CUP Model}\label{sec:requires-more}
As mentioned previously, at the beginning of the execution, each process $i$ has only access to $\mathit{PD}_i$ and $f$ in the CUP model.
In the Stellar model, it has only access to its slices. 
This section focuses on the following question: \textit{``Can slices be defined locally in the Stellar model using the information provided by the participant detectors to solve consensus in an unknown network with a known fault threshold?''}
or, equivalently, \textit{``Provided that each process $i$ has only access to $\mathit{PD}_i$ and $f$, can $i$ define its slices locally to form intertwined quorums that lead to a maximal consensus cluster?''}

We negatively answer those questions by first presenting two necessary properties that must be satisfied by the slices defined by each process $i$ using $\mathit{PD}_i$ and $f$.
Then, we show that two sets of processes $Q_1$ and $Q_2$ might be identified as quorums using $\mathtt{is\_quorum}$ (Algorithm~\ref{alg:is_quorum}) without satisfying $|Q_1\cap Q_2| > f$, i.e., it is impossible to ensure the formation of intertwined quorums if each process $i$ defines its slices locally using just $\mathit{PD}_i$ and $f$.
In the following, we present such necessary properties as lemmas.

\begin{lemma}\label{lem:slice-set}
    Provided that the slices of each process $i$ are defined locally using $\mathit{PD}_i$ and $f$, every slice $S$ of $i$ is a subset of $\mathit{PD}_i$, i.e., $\forall i \in \Pi, \forall S \in \mathcal{S}_{i}: S \subseteq \mathit{PD}_i$.
\end{lemma}
\begin{IEEEproof}
    Initially, each process $i$ only knows $\mathit{PD}_i$ and $f$. 
    Therefore, it can only define slices using processes contained in $\mathit{PD}_i$.
\end{IEEEproof}

\begin{lemma}\label{lem:available-slice}
    Each correct process $i$ must have at least one slice composed entirely of correct processes to solve consensus in Stellar. 
    Formally, let $\mathcal{B}_i$ be equal to $\{\forall B \subset \mathit{PD}_i : |B| \leq f\}$, then $\forall B \in \mathcal{B}_i$, $\exists S \in \mathcal{S}_i : S \cap B = \emptyset$.
\end{lemma}
\begin{IEEEproof}
    For the sake of contradiction, assume that $i$ does not have any slice composed entirely of correct processes.
    That is, each slice of $i$ has at least one faulty process.
    Since faulty processes can stay silent during an execution of a consensus instance, $i$ might not be able to make progress.
    Therefore, $i$ might not be able to solve consensus, which is a contradiction.
\end{IEEEproof}

\begin{theorem}\label{thm:safety-violation}
    If each process $i$ defines its slices locally using $\mathit{PD}_i$ and $f$, processes might violate the quorum intersection property.
\end{theorem}
\begin{IEEEproof}
    We prove this theorem by showing a counter-example.
    Consider $G_{\mathit{di}}$ as the one depicted in Fig. \ref{fig:31-osr}.
    This graph represents a $3$-OSR PD (see Definition~\ref{def:osr}), with $V_{\mathit{sink}} = \{1,2,3,4\}$, which provides enough knowledge for solving consensus with $f=1$. 
    Notice that whether the faulty process is a sink member or not, there are at least $2f+1=3$ correct sink members, there are at least $f+1=2$ node disjoint paths from any correct non-sink member to any correct sink member, and there are at least $f+1=2$ node disjoint paths from any correct sink member to another correct sink member.
    
    We can define the set of slices of each process $i$ as all subsets of $\mathit{PD}_i$ with size $|\mathit{PD}_i|-1$.
    In this way, we can ensure that each slice of $i$ is a subset of $\mathit{PD}_i$ (Lemma~\ref{lem:slice-set}) and $i$ has at least one slice composed entirely of correct processes (Lemma~\ref{lem:available-slice}).
    Set $Q_{1} = \{ 5,6,7 \}$ is a quorum because every process $j \in Q_{1}$ has a slice inside $Q_{1}$.
    Likewise, $Q_{2} = \{ 1,2,3,4 \}$ is also a quorum. 
    Since $Q_{1} \cap Q_{2} = \emptyset$, the quorum intersection property is violated.
    
    \begin{figure}[t!]
        \begin{minipage}{0.6\linewidth}
            \includegraphics[scale=0.17]{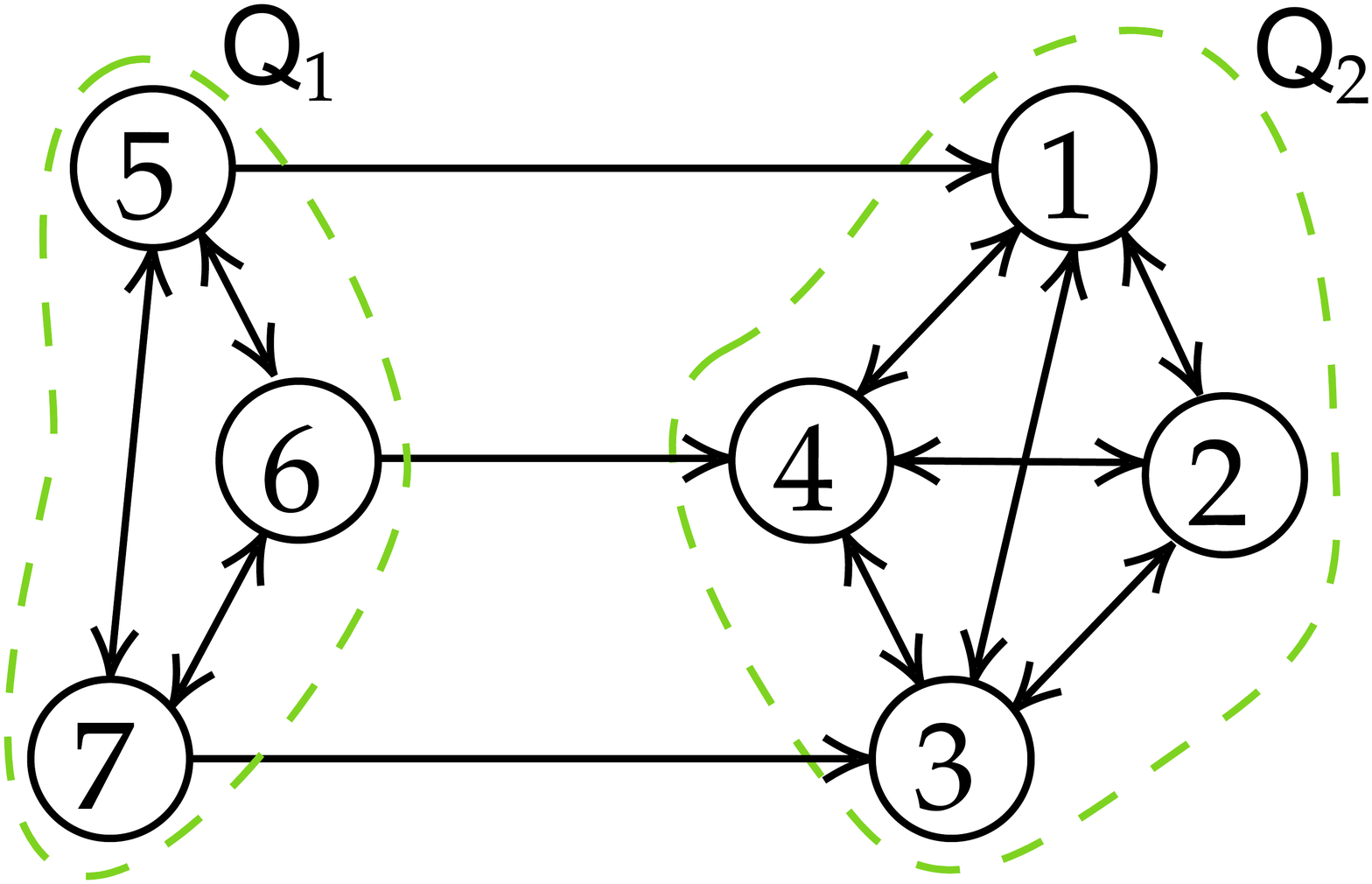}
        \end{minipage}
        \hspace{-0.2cm}
        \begin{minipage}{0.4\linewidth}
            \textbf{Processes' knowledge:}\\
            \hspace{3mm}$\mathit{PD}_1 = \{ 2,3,4 \}$ \\
            \hspace{3mm}$\mathit{PD}_2 = \{ 1,3,4 \}$ \\
            \hspace{3mm}$\mathit{PD}_3 = \{ 1,2,4 \}$ \\
            \hspace{3mm}$\mathit{PD}_4 = \{ 1,2,3 \}$ \\
            \hspace{3mm}$\mathit{PD}_5 = \{ 1,6,7\}$ \\
            \hspace{3mm}$\mathit{PD}_6 = \{ 4,5,7\}$ \\
            \hspace{3mm}$\mathit{PD}_7 = \{ 3, 5, 6 \}$
        \end{minipage}
        \caption{
            A knowledge connectivity graph satisfying $3$-OSR PD definition. 
            The dashed areas are two quorums, each formed by locally defined slices using PD and $f$.
        }
        \label{fig:31-osr}
    \end{figure}
\end{IEEEproof}

\begin{corollary}
    Stellar cannot solve Byzantine consensus with the minimal knowledge connectivity requirement of consensus.
\end{corollary}

    \section{Defining Slices Using Sink Detector}\label{sec:defining-slices}

We showed that Byzantine consensus might not be solved in the Stellar model if each process $i$ locally defines its slices using $\mathit{PD}_i$ and $f$.
The major problem with this approach is that in $k$-OSR, sink members might form quorums that do not intersect quorums formed by non-sink members.
In more detail, the knowledge connectivity graph obtained from PDs might be directed, and sink members do not know initially about non-sink members. 
Hence, a quorum formed by sink members contains only themselves.
On the other hand, non-sink members can form quorums without including sink members (see Theorem \ref{thm:safety-violation}). 

This problem can be solved by making non-sink members include the sink members in their slices.
Notice that the reverse is not possible since sink members do not have knowledge about non-sink members. 
This section introduces an oracle called \emph{sink detector} through which processes can discover the members of the sink component of a $k$-OSR knowledge connectivity graph.

\begin{definition}[Sink detector]\label{def:sink-detector}
    The Sink Detector (SD) is an oracle that provides an operation $\mathtt{get\_sink}$.
    Each process $i$ must provide $\mathit{PD}_i$ and $f$ as input to $\mathtt{get\_sink}$, which satisfies the following properties:
    \begin{itemize}
        \item If $i\in V_{\mathit{sink}}$, it returns $\langle \mathit{true}, V \rangle$ to $i$, where $V = V_{\mathit{sink}}$, and
        \item If $i \notin V_{\mathit{sink}}$, it returns $\langle \mathit{false}, V \rangle$, where $V \subseteq V_{\mathit{sink}}$ contains at least $f+1$ correct members of $V_{\mathit{sink}}$. 
    \end{itemize}
\end{definition}
It is important to remark that given a tuple $\langle *, V \rangle$ returned by $\mathtt{get\_sink}$, $V$ might contain faulty processes.  

\vspace{1em}
\noindent\textbf{Defining slices using SD.} 
Each process $i$ can get its slices by executing function $\mathtt{build\_slices}$, defined in Algorithm~\ref{alg:def-slices}.
This function defines slices using SD.
In further detail, first, $\mathtt{build\_slices}$ obtains the sink members by calling $\mathtt{get\_sink}$ (line \ref{line:get-sink}), which is based on $\mathit{PD}_i$ and $f$.
Then, it defines slices based on whether $i$ is a sink member or not, as follows:
\begin{itemize}
    \item If $i\in V_{\mathit{sink}}$, $\mathcal{S}_i$ contains all subsets of $V$ with size $\ceil{(|V|+f+1)/2}$ (line 3).
    
    \item If $i\notin V_{\mathit{sink}}$, $\mathcal{S}_i$ contains all subsets of $V$ with size $f+1$ (line 5).
\end{itemize}

\begin{algorithm}[t!]
    \caption{Building slices -- code of process $i$.}
    \label{alg:def-slices}
    \begin{algorithmic}[1]
        \STATEx{\hspace{-1.67em}\textbf{function} $\mathtt{build\_slices}(\mathit{PD}_i, f)$}
            \STATE{$\langle \mathit{flag}, V \rangle \leftarrow \mathtt{get\_sink}(\mathit{PD}_i, f)$}\label{line:get-sink}
            \NoThen
            \IF{$\mathit{flag} = \mathit{true}$}
                \STATE{$\mathcal{S}_i \leftarrow$ all subsets of $V$ with size $\ceil{(|V|+f+1)/2}$}
            \ELSE
                \STATE{$\mathcal{S}_i \leftarrow$ all subsets of $V$ with size $f+1$}
            \ENDIF
            \RETURN{$\mathcal{S}_i$}
    \end{algorithmic}
\end{algorithm}

The main idea behind defining slices this way is to enable us to define a lower bound for the size of quorums.
Recall that a set $Q$ is a quorum if each member of $Q$ has a slice contained within $Q$. 
Since slices defined using Algorithm~\ref{alg:def-slices} differ based on whether a process is inside or outside of the sink, quorums of the sink members will be different from quorums of non-sink members:
\begin{itemize}
    \item \textit{Quorums formed by sink members:}
    Consider any correct process $i \in V_{\mathit{sink}}$ and its respective quorum $Q_{i}$.
    Since $i$'s slices contain only sink members of size $\ceil{(|V_{\mathit{sink}}|+f+1)/2}$, $Q_i$ contains only sink members.
    Furthermore, $Q_i$'s size is greater than or equal to $\ceil{(|V_{\mathit{sink}}|+f+1)/2}$.
    \item \textit{Quorums formed by non-sink members:}
    Consider any correct non-sink member $j$ and its respective quorum $Q_j$.
    From the definition of quorums, process $j$ must have a slice $S$ contained within $Q_j$.
    Since each slice of $j$ contains $f+1$ sink members (lines 4-5 of Algorithm~\ref{alg:def-slices}), there is at least one correct sink member among them.
    Consequently, $S$ must contain a correct sink member $v$.
    Since each slice of $v$ has size $\ceil{(|V_{\mathit{sink}}|+f+1)/2}$ and $Q_j$ must contain a slice of $v$, $Q_j$'s size is greater than or equal to $\ceil{(|V_{\mathit{sink}}|+f+1)/2}$.
\end{itemize}

\begin{figure}
    \centering
    \includegraphics[scale=0.5]{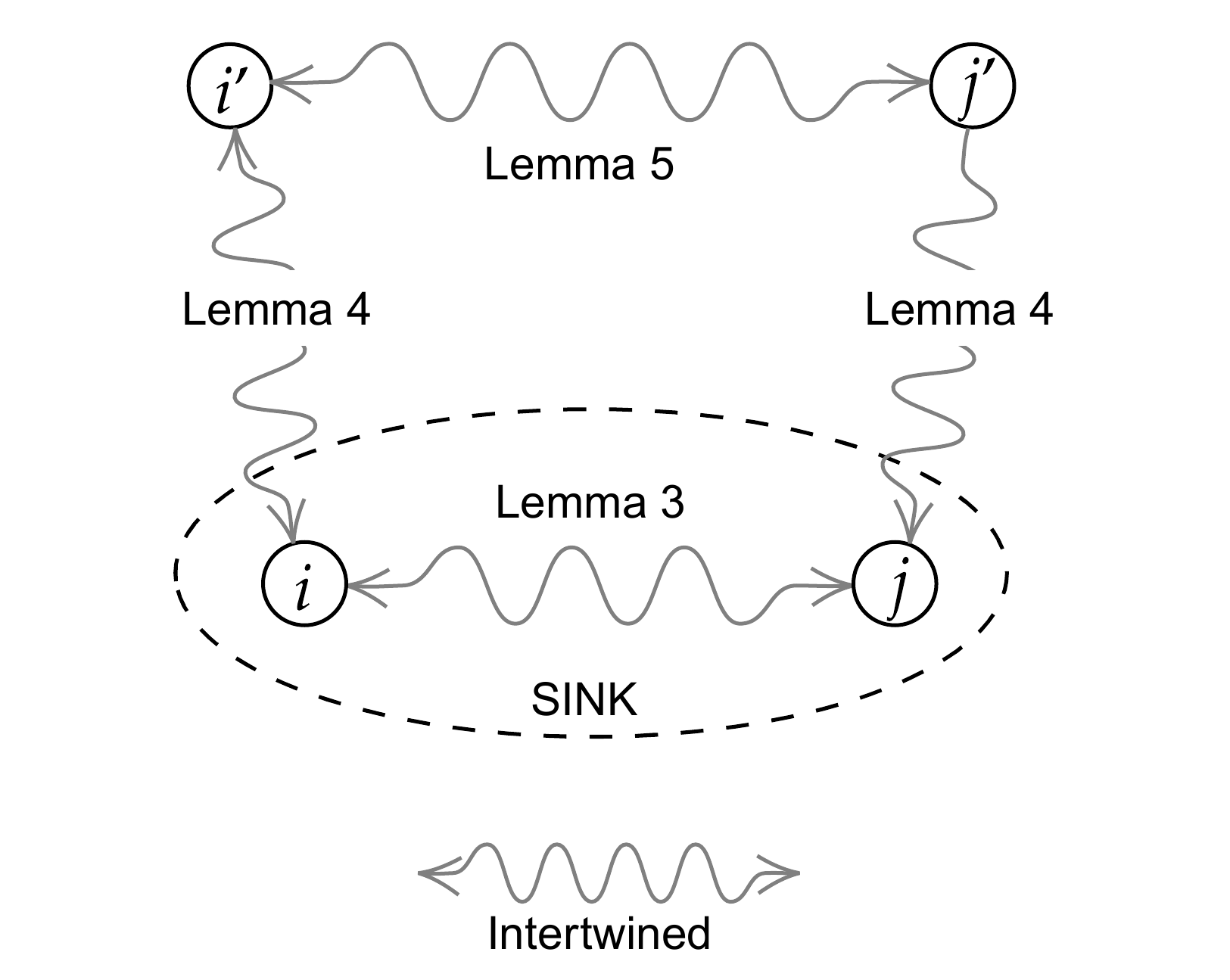}
    \caption{Illustration of the approach used to prove that each correct process is intertwined with another correct process.
    Processes $i, i', j$, and $j'$ are correct. 
    Processes $i$ and $j$ are sink members.}
    \label{fig:slices_sd_correctness_proof}
\end{figure}

\vspace{1em}
\noindent\textbf{Correctness proofs.}
By defining slices using Algorithm~\ref{alg:def-slices}, Stellar can solve consensus on the CUP model.
We prove this result by showing that every two correct processes are intertwined.
In further detail, we show how the sink and non-sink members are intertwined through three lemmas, as shown in Fig.~\ref{fig:slices_sd_correctness_proof}. 
At a high level, those lemmas show that every two correct processes' quorums intersect in at least $f+1$ sink members, i.e., every two correct processes are intertwined through the sink.
We put together these three lemmas in a theorem, proving the result.

The following lemma shows that every two correct sink members are intertwined due to their quorum intersections.

\begin{lemma}\label{lem:propert-q-in-sink}
    If slices of each sink member are defined using Algorithm~\ref{alg:def-slices}, then any two correct sink members are intertwined.
\end{lemma}
\begin{IEEEproof}
    Consider any two (possibly different) processes $i,j \in V_{\mathit{sink}}$.
    Let $Q_{i}$ be any quorum of $i$ and $Q_{j}$ be any quorum of $j$.
    To prove the lemma, we need to show that $|Q_i \cap Q_j| > f$.
    First, notice that $Q_i, Q_j \subseteq V_{\mathit{sink}}$.
    Since the sizes of both $Q_i$ and $Q_j$ are greater than or equal to $\ceil{(|V_{\mathit{sink}}|+f+1)/2}$, and there are at most $f$ faulty processes, it follows that $|Q_i \cap Q_j| > f$.
\end{IEEEproof}

The next lemma shows that every correct non-sink member is intertwined with every correct sink member because their quorums' intersections contain at least $f+1$ sink members.

\begin{lemma}\label{lem:propert-q-nonsink-sink}
    If slices of each process are defined using Algorithm~\ref{alg:def-slices}, then any correct sink member and any correct non-sink member are intertwined.
\end{lemma}
\begin{IEEEproof}
    Consider any correct non-sink member $i'$ (resp. correct sink member $i$) and its quorum $Q_{i'}$ (resp. $Q_i$).
    We need to show that $i'$ and $i$ are intertwined.
    Recall that any slice of $i'$ contains $f+1$ sink members, and according to  Definition~\ref{def:quorum}, $Q_{i'}$ contains slices of those sink members.
    Since any slice of each sink member has size $\ceil{(|V_{\mathit{sink}}|+f+1)/2}$, $Q_{i'}$ contains at least $\ceil{(|V_{\mathit{sink}}|+f+1)/2}$ members of the sink.
    On the other hand, quorum $Q_i$ contains at least $\ceil{(|V_{\mathit{sink}}|+f+1)/2}$ sink members, as $i$ is a sink member, and each of its slices contains at least $\ceil{(|V_{\mathit{sink}}|+f+1)/2}$ sink members.
    Therefore, $|Q_{i'} \cap Q_i|>f$, i.e., $i'$ and $i$ are intertwined.
\end{IEEEproof}

It remains to show that any two correct non-sink members are also intertwined. 
The following lemma formalizes it by showing that any correct non-sink member is intertwined with another correct non-sink member through sink members.

\begin{lemma}\label{lem:propert-q-nonsink-nonsink}
    If slices of each process are defined using Algorithm~\ref{alg:def-slices}, then any two correct non-sink members are intertwined.
\end{lemma}
\begin{IEEEproof}
    Consider any two correct non-sink members $i'$ and $j'$.
    Consider any quorum $Q_{i'}$ of $i'$ and any quorum  $Q_{j'}$ of $j'$.
    Let $i$ and $j$ be any two correct sink members such that $i \in Q_{i'}$ and $j \in Q_{j'}$.
    This is a valid assumption because quorums of non-sink members have at least $f+1$ sink members. 
    Since every member of $Q_{i'}$ has a slice contained within $Q_{i'}$, $Q_{i'}$ is also a quorum of $i$.
    Similarly, $Q_{j'}$ is also a quorum of $j$.
    Due to Lemma~\ref{lem:propert-q-in-sink}, $i$ and $j$ are intertwined, i.e., $|Q_{i} \cap Q_{j}| > f$.
	Accordingly, any quorum of $i'$ with any quorum of $j'$ has an intersection containing at least $f+1$ sink members, meaning that $i'$ and $j'$ are intertwined.
\end{IEEEproof}

\begin{theorem}\label{thm:cluster_intertwined}
    If slices of each process are defined using Algorithm~\ref{alg:def-slices}, then any two correct processes are intertwined.
\end{theorem}
\begin{IEEEproof}
The proof follows directly from Lemmata~\ref{lem:propert-q-in-sink}, \ref{lem:propert-q-nonsink-sink}, and \ref{lem:propert-q-nonsink-nonsink}.
\end{IEEEproof}

Recall that, from Definition~\ref{def:consensus_cluster}, Stellar requires a single maximal consensus cluster to solve consensus. Theorem~\ref{thm:cluster_intertwined} shows that when processes use PD, $f$, and SD to build slices, every two correct processes will be intertwined, which is one of two properties of consensus cluster. The following theorem proves that PD, $f$, and SD are sufficient to ensure the second property of consensus cluster, i.e., each correct process has at least one quorum composed entirely of correct processes.

\begin{theorem}\label{thm:cluster_intersection}
    Let $G_{di}$ be a knowledge connectivity graph with a sink component containing at least $2f + 1$ correct processes.
    If slices of each process in $G_{di}$ are defined using Algorithm~\ref{alg:def-slices}, then any correct process has at least one quorum composed entirely of correct processes.
\end{theorem}
\begin{IEEEproof}
    Let $G_{\mathit{sink}} = (V_{\mathit{sink}},E_{\mathit{sink}})$ be the sink component of $G_{di}$.
    We need to show that each correct process $i$ has at least one quorum composed entirely of correct processes.
    We consider two cases:
    \begin{enumerate}        
        \item $i \in V_{\mathit{sink}}$.
        Let $F_{\mathit{sink}}$ be the set containing all faulty processes of the sink. 
        Recall that each subset of $V_{\mathit{sink}}$ with size $\ceil{(|V_{\mathit{sink}}|+f+1)/2}$ is a slice of $i$.
        We first show that $i$ has at least one slice composed entirely of correct processes, i.e., $\exists S \in \mathcal{S}_i$ such that $S \cap F_{\mathit{sink}} = \emptyset$.
        To do so, we need to show that the following inequality holds:
        \begin{equation}\label{eq:40}
            |V_{\mathit{sink}}| \geq |F_{\mathit{sink}}| + \ceil{(|V_{\mathit{sink}}|+f+1)/2}
        \end{equation} 

        Since $|V_{\mathit{sink}}|$ and $|F_{\mathit{sink}}|$ are two natural numbers, Inequality \ref{eq:40} holds if and only if the following inequality holds:
        \begin{equation}\label{eq:40-1}
            |V_{\mathit{sink}}| \geq |F_{\mathit{sink}}| + (|V_{\mathit{sink}}|+f+1)/2
        \end{equation} 

        After simplifying Inequality \ref{eq:40-1}, we have:
        \begin{equation}\label{eq:41}
            |V_{\mathit{sink}}| \geq f+1  + 2|F_{\mathit{sink}}|
        \end{equation}

        Since $|F_{\mathit{sink}}| \leq f$, we have:
        \begin{align}\label{eq:43}
            2f+1 + |F_{\mathit{sink}}| \geq f+1  + 2|F_{\mathit{sink}}|
        \end{align}
        
        Since the sink component of $G_{\mathit{di}}$ contains at least $2f+1$ correct processes, we have:
        \begin{align}\label{eq:44}
            |V_{\mathit{sink}}| \geq 2f+1 + |F_{\mathit{sink}}|
        \end{align}

        By setting Inequalities \ref{eq:43} and \ref{eq:44} as the base and taking backward steps, it follows that Inequality \ref{eq:40} holds.
        Next, we show that a set $Q = S \cup \{ i \}$ is a quorum for $i$.
        To do so, we need to show that $\forall j \in Q$, $j$ has a slice in $Q$.
        Notice that $Q$ is composed of correct processes and its size is $\ceil{(|V_{\mathit{sink}}|+f+1)/2} +1$.
        Since any subset of $V_{\mathit{sink}}$ with size $\ceil{(|V_{\mathit{sink}}|+f+1)/2}$ is a slice for $j$, $j$ has a slice in $Q$.
        
        \item $i \notin V_{\mathit{sink}}$. 
        Due to Definition \ref{def:sink-detector}, $i$ has at least one slice $S'$ composed of correct sink members.
        From the first case, there must be a quorum $Q'$ composed of correct processes that is a quorum for each member of $S'$. 
        Notice that each member of $Q'' = Q' \cup \{ i \}$ has a slice in $Q''$, so $Q''$ is a quorum for $i$.
    \end{enumerate}
    The theorem holds since $i$ has at least one quorum composed only by correct processes in both cases.
\end{IEEEproof}

\begin{theorem}
    Let $G_{di}$ be a knowledge connectivity graph that is Byzantine-safe for $F$, and its sink component, $G_{\mathit{sink}} = (V_{\mathit{sink}}, E_{\mathit{sink}})$, contains at least $2f + 1$ correct processes.
    PD, $f$, and SD are sufficient to solve consensus in Stellar.
\end{theorem}
\begin{IEEEproof}
    We need to show that all correct processes form only one maximal consensus cluster using PD, $f$, and SD.
    According to Theorem~\ref{thm:cluster_intertwined}, every two correct processes are intertwined, which ensures the Quorum Intersection property.
    From Theorem~\ref{thm:cluster_intersection}, each correct process has a quorum composed entirely of correct processes, which ensures Quorum Availability.
    Since both properties of the consensus cluster are ensured, all correct processes form a consensus cluster $C$ using PD, $f$, and SD.
    Since $C$ contains all correct processes, it is maximal, proving the theorem.
\end{IEEEproof}

\begin{corollary}
    Having access to a sink detector, Stellar can solve consensus with the minimal knowledge connectivity requirement of Byzantine consensus.
\end{corollary}

    \section{Implementing the Sink Detector}\label{sec:implementing-sink}
This section presents an implementation of the sink detector using only the minimal knowledge required for solving consensus, i.e., the union of $\mathit{PD}_1, \mathit{PD}_2, \dots, \mathit{PD}_{|\Pi|}$ forms a $k$-OSR graph that is Byzantine-safe for $F$, and its sink component has at least $2f + 1$ correct processes.
This oracle discovers and returns members of the sink component.
When the $\mathtt{get\_sink}$ function is called, there are two ways to discover the sink.
The first way is to discover the sink directly.
However, it might be impossible, requiring the indirect discovery of the sink.
In the following, we elaborate on each way.

\vspace{1em}
\noindent\textbf{Discovering the sink directly.}
Each process $i$ calls $\mathtt{SINK}(\mathit{PD}_i,f)$ function, presented in \cite{alchieri_2018}, to discover the sink directly.
In a nutshell, $\mathtt{SINK}$ consists of three steps:
\begin{enumerate}
    \item It runs a kind of breadth-first search in $G_{\mathit{di}}$ to obtain the maximal set of processes that $i$ can reach and stores it in a variable $\mathit{known}_i$.
    Every sink member terminates this step; however, a non-sink member might not be able to terminate (to see the reason, see Section 4 of \cite{alchieri_2018}).
    \item After obtaining $\mathit{known}_i$, process $i$ sends $\mathit{known}_i$ to every process it knows.
    \item If $i$ receives at least $|\mathit{known}_i|-f$ messages with the same content as $\mathit{known}_i$, then $i$ is a sink member, and the algorithm terminates by returning $\langle \mathit{true}, V_{\mathit{sink}}\rangle$.
    Otherwise, if $i$ receives at least $f+1$ messages with different sets than $\mathit{known}_i$, it is a non-sink member. 
\end{enumerate}

The following lemma proves that $\mathtt{SINK}$ terminates at sink members by returning the sink members. See \cite{alchieri_2018} for the proof.
\begin{lemma}[Sink members -- $\mathtt{SINK}$ \cite{alchieri_2018}]\label{lem:sink:alchieri_2016}
    Function $\mathtt{SINK}$ executed by a correct process $i\in G_{\mathit{sink}}$ satisfies the following properties:
    \begin{itemize}
        \item \emph{Sink Termination}: $i$ terminates the execution, and
        \item \emph{Sink Accuracy}: $i$ returns $\langle \mathit{true}, V_{\mathit{sink}} \rangle$.
    \end{itemize}
\end{lemma}

Since non-sink members might not terminate the first step in the $\mathtt{SINK}$ function, they cannot use $\mathtt{SINK}$ to discover the sink directly.
Hence, in addition to executing $\mathtt{SINK}$, each process $i$ might need to discover the sink indirectly. 

\vspace{1em}
\noindent\textbf{Discovering the sink indirectly.}
Each process $i$ asks sink members to send the sink component to it.
If a sink member discovers the sink and receives $i$'s request, it sends the sink to $i$.
A primitive called \emph{reachable-reliable broadcast}, also presented in \cite{alchieri_2018}, is used by $i$ to communicate with sink members.
The primitive provides two operations:
\begin{itemize}
    \item $\mathtt{reachable\_bcast}(m,i)$ -- through which the process $i$ broadcasts message $m$ to all $f$-reachable processes from $i$ in $G_{\mathit{di}}$.
    \item $\mathtt{reachable\_deliver}(m,i)$ -- invoked by a receiver to deliver message $m$ sent by the process $i$.
\end{itemize}

This primitive is based on the notion of $f$-reachability.

\begin{definition}[$f$-reachability \cite{alchieri_2018}]
Consider a knowledge connectivity graph $G_{\mathit{di}}$ and let $f$ be the number of processes in $G_{\mathit{di}}$ that may fail. 
For any two processes $i, j \in G_{\mathit{di}}$, $j$ is $f$-reachable from $i$ in $G_{\mathit{di}}$ if there are at least $f +1$ node-disjoint paths from $i$ to $j$ in $G_{\mathit{di}}$ composed only by correct processes.
\end{definition}

The reachable-reliable broadcast should satisfy the following properties:

\begin{itemize}
    \item \emph{RB\_Validity}: If a correct process $i$ invokes $\mathtt{reachable\_bcast}(m,i)$ then (i) some correct process $j$,
    $f$-reachable from $i$ in $G_{\mathit{di}}$, eventually invokes $\mathtt{reachable\_deliver}(m,i)$ or (ii) there is no correct process $f$-reachable from $i$ in $G_{\mathit{di}}$.

    \item \emph{RB\_Integrity}: For any message $m$, if a correct process $j$ invokes 
    $\mathtt{reachable\_deliver}(m,i)$ then process $i$ has invoked 
    $\mathtt{reachable\_bcast}(m,i)$.

    \item \emph{RB\_Agreement}: If a correct process $j$ invokes $\mathtt{reachable\_deliver}(m,i)$, where $m$ was sent by a correct process $i$ that invoked $\mathtt{reachable\_bcast}(m,i)$, then all correct processes $f$-reachable from $i$ in $G_{\mathit{di}}$ invoke $\mathtt{reachable\_deliver}(m,i)$.    
\end{itemize}

This primitive was implemented in asynchronous systems, and it was shown that all sink members are $f$-reachable from any process in $G_{\mathit{di}}$ \cite{alchieri_2018}.
Therefore, if any process broadcasts a message using $\mathtt{reachable\_bcast}$, all correct sink members deliver the message using $\mathtt{reachable\_deliver}$.
Accordingly, any non-sink member will discover the sink members with the help of sink members.

\vspace{1em}
\noindent\textbf{Description of $\mathtt{get\_sink}$ (Algorithm \ref{alg:sd}).} 
When $\mathtt{get\_sink}$ is called, each process $i$ examines whether it has discovered the sink.
If it is not the case, it asks processes to send the sink to it by broadcasting a message tagged with $\mathtt{GET\_SINK}$ (line \ref{line:reachable-bcast}).
By delivering a message tagged with $\mathtt{GET\_SINK}$ sent by a process $j$, $i$ adds $j$ to the set $\mathit{asked}$ (line \ref{line:reachable-deliver}).
Then, $i$ executes $\mathtt{SINK}(\mathit{PD}_i,f)$.
If $i \in V_{\mathit{sink}}$, $\mathtt{SINK}$ terminates by returning $\langle \mathit{true}, V_{\mathit{sink}} \rangle$, and $i$ sends $V_{\mathit{sink}}$ to every member of $\mathit{asked}$ (lines \ref{line:sink-send}-\ref{line:sink-end}). 
Otherwise, $i$ must wait until the sink members send the sink to it.
By receiving a value $v$ from any other process, $i$ adds $v$ to the list $\mathit{values}$.
If there is a value $v$ that is repeated more than $f$ times in $\mathit{values}$, $i$ selects $v$ as the sink (lines \ref{line:wait-sink}-\ref{line:assign-sink}).
As soon as $i$ finds the sink, it returns the sink.

\begin{algorithm}[t!]
    \caption{SD code of process $i$.}
    \label{alg:sd}
    \begin{algorithmic}[1]
        \STATEx{\hspace{-1.67em}\textbf{variable}}
            \STATE{$\mathit{sink} \leftarrow \emptyset$ {\color{gray!90} $\qquad/*$ a set that will be filled with $V_{\mathit{sink}}$ eventually $*/$}}
            \STATE{$\mathit{asked} \leftarrow \emptyset$ {\color{gray!90} $\qquad/*$ a set containing the ids of processes that asked $i$ about the sink $*/$}}
            \STATE{$\mathit{values} \leftarrow \emptyset$ {\color{gray!90} $\qquad/*$ a list containing the values returned by other processes $*/$}}

        \vspace{0.5em}
        \STATEx{\hspace{-1.67em}\textbf{function} $\mathtt{get\_sink}(\mathit{PD}_i, f)$}
            \NoThen
            \IF{$\mathit{sink} = \emptyset$}
                \STATE{$\mathtt{reachable\_bcast}(\mathtt{GET\_SINK}, i)$}\label{line:reachable-bcast}
                \STATE{\textbf{fork} $\mathtt{wait\_sink}()$}
                \NoThen
                \IF{$\langle \mathit{true}, V_{\mathit{sink}} \rangle = $ $\mathtt{SINK}(\mathit{PD}_i, f)$ {\color{gray!90} $\qquad/*$ executing the $\mathtt{SINK}$ algorithm from \cite{alchieri_2018} $*/$}}
                    \STATE{$\mathit{sink} \leftarrow V_{\mathit{sink}}$}
                    \STATE{\textbf{fork} $\mathtt{send\_sink}()$}
                \ENDIF
            \ENDIF
            \STATE{\textbf{wait until} $\mathit{sink} \neq \emptyset$}
            \IF{$i \in \mathit{sink}$}
                \STATE{\textbf{return} $\langle \mathit{true}, \mathit{sink}\rangle$}
            \ELSE
                \STATE{\textbf{return} $\langle \mathit{false}, \mathit{sink}\rangle$}\label{line:return-sink}
            \ENDIF

        \vspace{0.5em}
        \STATEx{\hspace{-1.67em}\textbf{function} $\mathtt{wait\_sink}()$}
            \STATE{\textbf{wait until} there is a value $v$ that is repeated more than $f$ times in $\mathit{values}$}\label{line:wait-sink}
            \STATE{$\mathit{sink} = v$}\label{line:assign-sink}\label{line:receive-sink}

        \vspace{0.5em}
        \STATEx{\hspace{-1.67em}\textbf{upon} $\mathtt{reachable\_deliver}(\mathtt{GET\_SINK}, j)$}
            \STATE{$\mathit{asked} \leftarrow \mathit{asked} \cup \{ j \}$}\label{line:reachable-deliver}

        \vspace{0.5em}
        \STATEx{\hspace{-1.67em}\textbf{function} $\mathtt{send\_sink}()$}
            \LOOP\label{line:sink-send} 
                \NoThen
                \IF{there is a process $j \in \mathit{asked}$}
                    \STATE{\textbf{send} $\langle \mathtt{SINK}, \mathit{sink} \rangle$ to $j$}
                    \STATE{$\mathit{asked} \leftarrow \mathit{asked} \setminus \{ j \}$}\label{line:sink-end}
                \ENDIF
            \ENDLOOP

        \vspace{0.5em}
        \STATEx{\hspace{-1.67em}\textbf{upon receiving} $\langle \mathtt{SINK}, V \rangle$}
            \STATE{$\mathit{values} \leftarrow \mathit{values} \cup \{V\}$}
    \end{algorithmic}
\end{algorithm}

\begin{theorem}
    If a correct process calls $\mathtt{get\_sink}$ (Algorithm \ref{alg:sd}), it will eventually receive $V_{\mathit{sink}}$.
\end{theorem}
\begin{IEEEproof}
    Let $i$ be a correct process.
    We need to consider two cases:
    \begin{enumerate}
        \item $i\in V_{\mathit{sink}}$.
        Since the invocation of $\mathtt{SINK}$ terminates by returning $\langle \mathit{true}, V_{\mathit{sink}} \rangle$ to $i$ according to Lemma \ref{lem:sink:alchieri_2016}, the theorem holds for this case.
        
        \item $i\notin V_{\mathit{sink}}$.
        Notice that members of the sink are $f$-reachable from $i$, so every correct sink member will receive $\langle \mathtt{GET\_SINK}, i \rangle$.
        Since $\mathtt{SINK}$ terminates in every correct process $j \in V_{\mathit{sink}}$, $j$ will obtain $V_{\mathit{sink}}$ and can send it to $i$.
        Since there are at least $2f+1$ correct processes inside the sink, $i$ will receive more than $f$ values that are equal to $V_{\mathit{sink}}$.
        It follows that $i$ can eventually learn $V_{\mathit{sink}}$.
    \end{enumerate}
\end{IEEEproof}

Each process uses Algorithm \ref{alg:sd} to obtain the sink members, which are used in Algorithm \ref{alg:def-slices} to define its slices forming a consensus cluster.

    \section{Conclusion}\label{sec:conclusion}

We studied the required knowledge for Stellar to solve consensus in open systems.
We showed that it is impossible to ensure the formation of a consensus cluster when each participant defines its slices locally using the fault threshold and a list of participants defined by the minimal knowledge connectivity graph required for solving Byzantine consensus.
We also proposed an oracle -- the sink detector -- that provides the information required by each participant to define slices that lead to the formation of a consensus cluster.

These results imply that, differently from the BFT-CUP protocol~\cite{alchieri_2018}, Stellar cannot solve consensus when processes have only the minimal required knowledge about the system.
Further, to make Stellar solve consensus in such conditions, processes need to run some distributed knowledge-increasing protocol before building their slices.
An interesting question for future work is if the BFT-CUP approach can be used for implementing a permissionless blockchain.

    \section*{Acknowledgments}
        We thank the ICDCS'23 anonymous reviewers for providing constructive comments to improve this paper.
        This work was supported by FCT through a Ph.D. scholarship (2020.04412.BD), the SMaRtChain project (2022.08431.PTDC), and the LASIGE Research Unit (UIDB/00408/2020 and UIDP/00408/2020), and by European Commission through the VEDLIoT project (H2020 957197).
    \bibliographystyle{IEEEtran}
    \bibliography{ref.bib}

\begin{thebibliography}{10}
\providecommand{\url}[1]{#1}
\csname url@samestyle\endcsname
\providecommand{\newblock}{\relax}
\providecommand{\bibinfo}[2]{#2}
\providecommand{\BIBentrySTDinterwordspacing}{\spaceskip=0pt\relax}
\providecommand{\BIBentryALTinterwordstretchfactor}{4}
\providecommand{\BIBentryALTinterwordspacing}{\spaceskip=\fontdimen2\font plus
\BIBentryALTinterwordstretchfactor\fontdimen3\font minus
  \fontdimen4\font\relax}
\providecommand{\BIBforeignlanguage}[2]{{%
\expandafter\ifx\csname l@#1\endcsname\relax
\typeout{** WARNING: IEEEtran.bst: No hyphenation pattern has been}%
\typeout{** loaded for the language `#1'. Using the pattern for}%
\typeout{** the default language instead.}%
\else
\language=\csname l@#1\endcsname
\fi
#2}}
\providecommand{\BIBdecl}{\relax}
\BIBdecl

\bibitem{lamport_1982}
L.~Lamport, R.~Shostak, and M.~Pease, ``{The Byzantine Generals Problem},''
  \emph{ACM Trans. Program. Lang. Syst.}, vol.~4, no.~3, 1982.

\bibitem{schneider_1990}
F.~B. Schneider, ``{Implementing Fault-Tolerant Services Using the State
  Machine Approach: A Tutorial},'' \emph{ACM Computing Surveys}, vol.~22,
  no.~4, 1990.

\bibitem{castro_1999}
M.~Castro and B.~Liskov, ``{Practical Byzantine Fault Tolerance},'' in
  \emph{Symposium on Operating Systems Design and Implementation}, 1999.

\bibitem{dwork_1988}
C.~Dwork, N.~Lynch, and L.~Stockmeyer, ``{Consensus in the Presence of Partial
  Synchrony},'' \emph{Journal of the ACM}, vol.~35, no.~2, 1988.

\bibitem{lamport_1998}
L.~Lamport, ``The part-time parliament,'' \emph{ACM Transactions on Computer
  Systems}, vol.~16, no.~2, 1998.

\bibitem{ongaro_2014}
D.~Ongaro and J.~Ousterhout, ``{In Search of an Understandable Consensus
  Algorithm},'' in \emph{USENIX Annual Technical Conference}, 2014.

\bibitem{nakamoto_2008}
S.~Nakamoto, ``{Bitcoin: A Peer-to-Peer Electronic Cash System},'' 2008.

\bibitem{vukolic_2015}
M.~Vukoli{\'c}, ``{The Quest for Scalable Blockchain Fabric: Proof-of-Work vs.
  BFT Replication},'' in \emph{International Workshop on Open Problems in
  Network Security}, 2015.

\bibitem{abraham_2016}
I.~Abraham, D.~Malkhi, K.~Nayak, L.~Ren, and A.~Spiegelman, ``{Solida: A
  Blockchain Protocol Based on Reconfigurable Byzantine Consensus},'' in
  \emph{International Conference on Principles of Distributed Systems}, 2018.

\bibitem{decker_2016}
C.~Decker, J.~Seidel, and R.~Wattenhofer, ``{Bitcoin Meets Strong
  Consistency},'' in \emph{International Conference on Distributed Computing
  and Networking}, 2016.

\bibitem{pass_2017_hybrid_consensus}
R.~Pass and E.~Shi, ``{Hybrid Consensus: Efficient Consensus in the
  Permissionless Model},'' in \emph{International Symposium on Distributed
  Computing}, 2017.

\bibitem{garcia_2018}
{\'A}.~Garc{\'\i}a-P{\'e}rez and A.~Gotsman, ``{Federated Byzantine Quorum
  Systems},'' in \emph{International Conference on Principles of Distributed
  Systems}, 2018.

\bibitem{lokhava_2019}
M.~Lokhava, G.~Losa, D.~Mazi{\`e}res, G.~Hoare, N.~Barry, E.~Gafni, J.~Jove,
  R.~Malinowsky, and J.~McCaleb, ``{Fast and Secure Global Payments with
  Stellar},'' in \emph{ACM Symposium on Operating Systems Principles}, 2019.

\bibitem{schawartz_2014}
D.~Schwartz, N.~Youngs, A.~Britto \emph{et~al.}, ``{The Ripple Protocol
  Consensus Algorithm},''
  \url{https://ripple.com/files/ripple\_consensus\_whitepaper.pdf}, 2014.

\bibitem{mazieres_2015}
D.~Mazieres, ``{The Stellar Consensus Protocol: A Federated Model for
  Internet-level Consensus},''
  \url{https://stellar.org/papers/stellar-consensus-protocol.pdf}, 2015.

\bibitem{losa_2019}
G.~Losa, E.~Gafni, and D.~Mazi{\`e}res, ``{Stellar Consensus by
  Instantiation},'' in \emph{International Symposium on Distributed Computing},
  2019.

\bibitem{alchieri_2018}
E.~A.~P. Alchieri, A.~Bessani, F.~Greve, and J.~da~Silva~Fraga, ``{Knowledge
  Connectivity Requirements for Solving Byzantine Consensus with Unknown
  Participants},'' \emph{IEEE Transactions on Dependable and Secure Computing},
  vol.~15, no.~2, 2018.

\bibitem{cavin_2004}
D.~Cavin, Y.~Sasson, and A.~Schiper, ``{Consensus with Unknown Participants or
  Fundamental Self-Organization},'' in \emph{International Conference on Ad-Hoc
  Networks and Wireless}, 2004.

\bibitem{cavin_2005}
------, ``{Reaching Agreement with Unknown Participants in Mobile
  Self-Organized Networks in Spite of Process Crashes},'' EPFL - LSR, Tech.
  Rep., 2005.

\bibitem{greve_2007}
F.~Greve and S.~Tixeuil, ``{Knowledge Connectivity vs. Synchrony Requirements
  for Fault-Tolerant Agreement in Unknown Networks},'' in \emph{Annual
  IEEE/IFIP International Conference on Dependable Systems and Networks}, 2007.

\bibitem{damgard_2007}
I.~Damg{\aa}rd, Y.~Desmedt, M.~Fitzi, and J.~B. Nielsen, ``{Secure Protocols
  with Asymmetric Trust},'' in \emph{International Conference on the Theory and
  Application of Cryptology and Information Security}, 2007.

\bibitem{chase_2018}
B.~Chase and E.~MacBrough, ``Analysis of the {XRP} ledger consensus protocol,''
  \emph{arXiv preprint arXiv:1802.07242}, 2018.

\bibitem{amores_2020}
I.~Amores-Sesar, C.~Cachin, and J.~Mi{\'c}i{\'c}, ``Security analysis of ripple
  consensus,'' in \emph{International Conference On Principles Of Distributed
  Systems}, 2020.

\bibitem{malkhi_1998}
D.~Malkhi and M.~Reiter, ``{Byzantine Quorum Systems},'' \emph{Distributed
  Computing}, vol.~11, no.~4, 1998.

\bibitem{garcia_2019}
Álvaro Garc{\'\i}a-P{\'e}rez and M.~A. Schett, ``{Deconstructing Stellar
  Consensus},'' in \emph{International Conference On Principles Of Distributed
  Systems}, 2020.

\bibitem{cachin_2019}
C.~Cachin and B.~Tackmann, ``Asymmetric distributed trust,'' in
  \emph{International Conference On Principles Of Distributed Systems}, 2019.

\bibitem{cachin_2022}
C.~Cachin, G.~Losa, and L.~Zanolini, ``{Quorum Systems in Permissionless
  Network},'' in \emph{International Conference On Principles Of Distributed
  Systems}, 2022.

\bibitem{chandra_1996}
T.~D. Chandra and S.~Toueg, ``{Unreliable Failure Detectors for Reliable
  Distributed Systems},'' \emph{Journal of the ACM}, vol.~43, no.~2, 1996.

\bibitem{alchieri_2008}
E.~A. Alchieri, A.~N. Bessani, J.~d. Silva~Fraga, and F.~Greve, ``{Byzantine
  Consensus with Unknown Participants},'' in \emph{International Conference On
  Principles Of Distributed Systems}, 2008.

\bibitem{khanchandani_2021}
P.~Khanchandani and R.~Wattenhofer, ``{Byzantine Agreement with Unknown
  Participants and Failures},'' in \emph{2021 IEEE International Parallel and
  Distributed Processing Symposium}, 2021.

\bibitem{momose_2022}
A.~Momose and L.~Ren, ``{Constant Latency in Sleepy Consensus},'' in
  \emph{Proceedings of the 2022 ACM SIGSAC Conference on Computer and
  Communications Security}, 2022.

\bibitem{pass_2017}
R.~Pass and E.~Shi, ``{The Sleepy Model of Consensus},'' in \emph{International
  Conference on the Theory and Application of Cryptology and Information
  Security}, 2017.

\bibitem{biely_2012}
M.~Biely, P.~Robinson, and U.~Schmid, ``{Agreement in Directed Dynamic
  Networks},'' in \emph{International Colloquium on Structural Information and
  Communication Complexity}, 2012.

\bibitem{vaidya_2012}
N.~H. Vaidya, L.~Tseng, and G.~Liang, ``{Iterative Approximate Byzantine
  Consensus in Arbitrary Directed Graphs},'' in \emph{ACM Symposium on
  Principles of Distributed Computing}, 2012.

\bibitem{biely_2018}
M.~Biely, P.~Robinson, U.~Schmid, M.~Schwarz, and K.~Winkler, ``{Gracefully
  Degrading Consensus and k-set Agreement in Directed Dynamic Networks},''
  \emph{Theoretical Computer Science}, vol. 726, 2018.

\bibitem{tseng_2015}
L.~Tseng and N.~H. Vaidya, ``{Fault-Tolerant Consensus in Directed Graphs},''
  in \emph{ACM Symposium on Principles of Distributed Computing}, 2015.

\bibitem{douceur_2002}
J.~R. Douceur, ``{The Sybil Attack},'' in \emph{International Workshop on
  Peer-to-Peer Systems}, 2002.

\bibitem{fischer_1983}
M.~J. Fischer, ``The consensus problem in unreliable distributed systems (a
  brief survey),'' in \emph{International Conference on Fundamentals of
  Computation Theory}, 1983.

\bibitem{malkhi_1997}
D.~Malkhi and M.~Reiter, ``{Byzantine Quorum Systems},'' in \emph{Annual
  Symposium on Theory of Computing}, 1997.

\end{thebibliography}
\end{document}